\documentclass[a4paper]{article}
\usepackage{mathptmx}
\usepackage{graphicx}
\usepackage{wrapfig}
\usepackage{rotating}
\usepackage{subfigure}
\usepackage{amssymb}
\usepackage{longtable}
\usepackage[top=2.5cm, bottom=2.5cm, left=2.5cm, right=2.5cm]{geometry} 
\usepackage{placeins}
\usepackage[footnotesize,bf]{caption}
\usepackage{amssymb}
\usepackage{float}
\usepackage{natbib}
\usepackage{hyperref}
\hypersetup{
    colorlinks=true,
    linkcolor=blue,
    filecolor=magenta,      
    urlcolor=blue,
}

\begin{document}

\title{Comment on `The role of 3-D interactive visualization in blind surveys of HI in galaxies'}
\date{June 2015}
\author{Rhys Taylor \\Astronomical Institute of the Czech Academy of Sciences, Prague, Czech Republic\\ \small Contact : rhysyt@gmail.com}

\maketitle

\begin{abstract}

\noindent \cite{P15} recently reported on the state of the art for visualisation of \textsc{Hi} data cubes. I here briefly describe another program, \textsc{frelled}, specifically designed for dealing with \textsc{Hi} data. Unlike many 3D viewers, \textsc{frelled} can handle astronomical world coordinates, easily and interactively mask and label specific volumes within the data, overlay optical data from the SDSS, generate contour plots and renzograms, make basic spectral profile measurements via an interface with \textsc{miriad}, and can switch between viewing the data in 3D and 2D. The code is open source and can potentially be extended to include any astronomical function possible with Python, displaying the result in an interactive 3D environment.

\end{abstract}

\section{Introduction}
\cite{P15} (hereafter P15) have described in detail the need for a capable 3D viewer for analysis of \textsc{Hi} data cubes. They describe the abilities such a viewer must have and assess four such viewers in this context. Two of these are general purpose 3D data viewers, two are designed for medial data. They note :
``A recent development is the use of the open source software Blender for visualisation of astronomical data (\citealt{bkent}; \citealt{me14} hereafter T14), but this application is more suitable for data presentation than interactive data analysis''.

While Blender itself is not designed as a data viewer (it is primarily aimed at artists), it has many characteristics that make it eminently suitable as such. The most important of these for astronomy is the presence of a Python interpreter. This allows Blender to interface with any of the numerous astronomical routines available in Python. Coupled with its interactive 3D display (which most other Python interpreters lack) and other built-in features (I will describe some of these in more detail below), it becomes a very powerful tool for interactive 3D data analysis.

\begin{figure}
\begin{center}
\includegraphics[width=160mm]{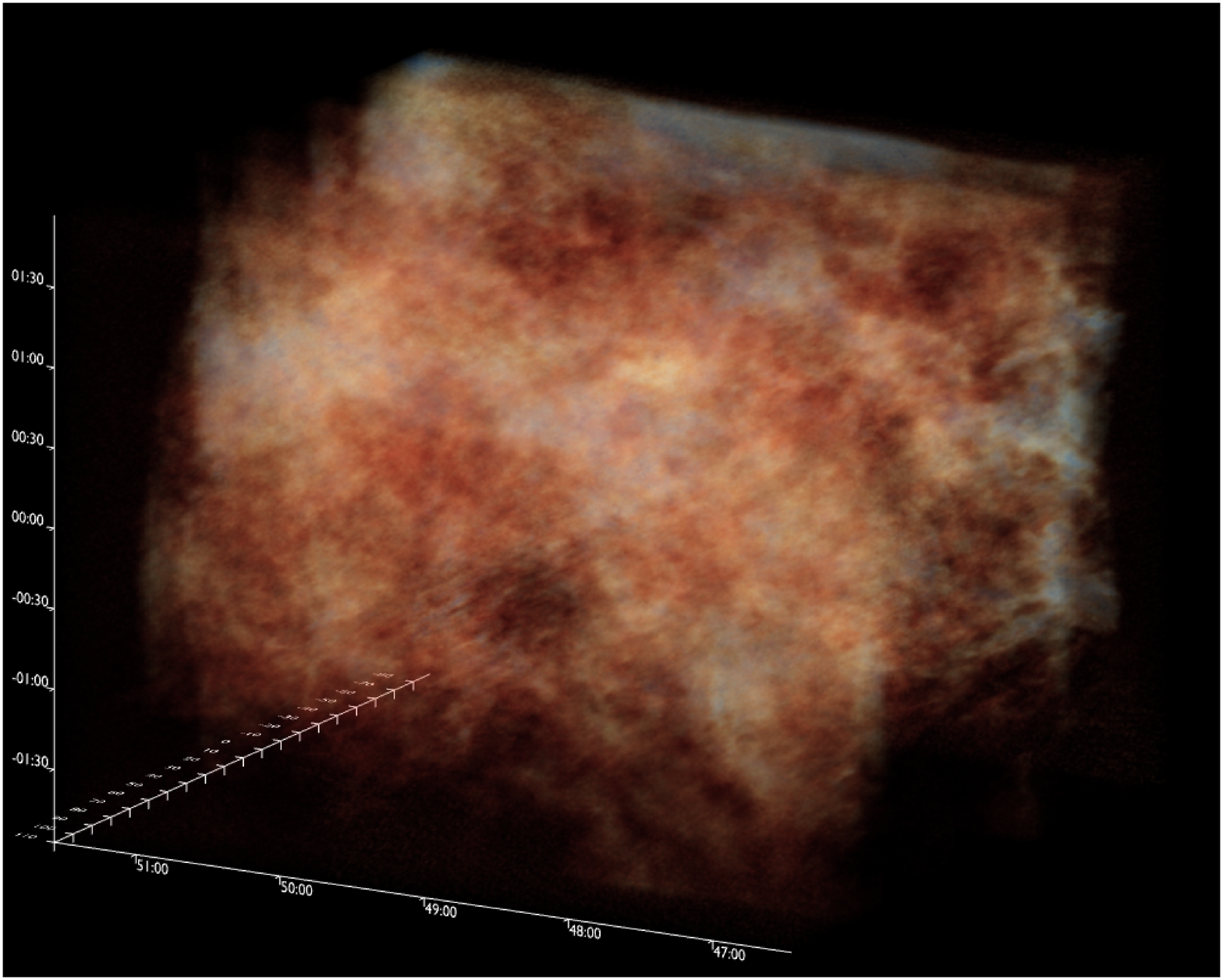}
\caption[VLAGPS]{Relatime volume render of a VLA Galactic Plane Survey (\citealt{stil}) data cube in \textsc{frelled}. A rotating movie of this figure can be seen at \href{https://www.youtube.com/watch?v=7r0UJGzLK8M}{this url}.}
\label{fig:VLAGPS}
\end{center}
\end{figure}

\textsc{Frelled} (Fits Realtime Explorer of Low Latency in Every Dimension) is a set of Python scripts which can be used to visualise 3D astronomical data inside Blender. An example of visualising a data set with \textsc{frelled} is shown in figure \ref{fig:VLAGPS}. Two principle reasons motivated the development of \textsc{frelled} : 1) Blender's powerful 3D display means we can leapfrog the development of the display routine itself; 2) The need to easily and interactively mask and analyse \textsc{Hi} data, described in P15 and T14, which is wholly lacking from other 3D astronomical software. \textsc{Frelled}, unlike the viewers described in P15, was thus explicitly designed with astronomical data in mind, particularly \textsc{Hi} data. A full description of \textsc{frelled} can be found on the \textsc{frelled} \href{http://frelled.wikia.com/wiki/FRELLED_Wikia}{wiki page}. Here I present a brief summary of what \textsc{frelled} can and could do and how it compares to other viewers.

\section{FRELLED in brief}
By using the realtime display capabilities of Blender, \textsc{frelled} is a GPU-based volume renderer. The methodology is as follows. Since Blender cannot view FITS data directly, the cube is ``sliced'' into a series of PNG images, each of a constant velocity, declination, or right ascension. These images are mapped onto a sequence of Blender's internal mesh objects (planes), which are separated by an amount equivalent to the dimensions of one pixel in the PNG images. The user controls the colour transfer function (before converting the cube into image files) and the transparency of the image planes.

\textsc{Frelled} is now developed on an HP Elite 7500 series desktop with an Intel i7-3770 quad core 3.4 GHz CPU, 16 GB RAM and a 4GB NVIDIA GeForce GT 640 GPU. Display of the FITS file in FRELLED depends mainly on the GPU while speed of operations that process or analyse the data are mainly CPU-dependent.

Although of course all features in \textsc{frelled} are ultimately enabled by Blender, in this rest of this document I will refer to Blender when dealing with a feature explicitly provided by the software with no Python code required, and \textsc{frelled} when referring to features which are enabled - or made simpler - by our scripts.

\section{Requirements for Visualisation of HI}
P15 have described the ideal features an \textsc{Hi} viewer. Here I describe which of these \textsc{frelled} already has, those which it does not currently possess but could, and those which it is not capable of achieving. Following P15 I divide the features into several areas : 1) Qualitative analysis features, which allow the user to visualise the data in different ways; 2) Quantitative analysis features; 3) Comparative analysis features, which allow the user to link their data to other sources; 4) Software and code features, e.g. user interface accessibility and documentation of the code.

I here take the term ``realtime'' to mean that any change is visible instantaneously as the user alters some parameter. An ``interactive'' feature must be not only be, obviously, controlled directly by the user, but must also occur on a timescale of a few seconds or less. Without this limitation almost any feature available in Python could be implemented in \textsc{frelled} (or Blender) ``interactively''.

\subsection{Qualitative analysis}
As noted by P15, any \textsc{Hi} viewer useful for large upcoming surveys with ASKAP, Meerkat, and APERTIF, must be capable of handling large data sets and high source densities. \textsc{Frelled} is not capable of dealing with the 4096$^{3}$ voxel cubes which will be returned by APERTIF, however it is certainly capable of displaying 600$^{3}$ voxel data sets even on current workstations. This is more than adequate for the typical $<$512$^{3}$ voxels for expected subset data cubes sent to users by automated algorithms for further analysis.

T14 demonstrated that \textsc{frelled} is capable of dealing with very high source densities, allowing a single user to mask and catalogue $\sim$300 sources within a single working day. In that data set, the sources occupied about 10\% of the volume, far higher than the expected 0.0001\% for SKA precursor surveys as described in P15 (though this will rise as automated source-finders return much smaller subset cubes containing possible sources). Thus, \textsc{frelled} is already suitable for source extraction from large data sets such as those from APERTIF.

\begin{figure}[h]
\begin{center}
\includegraphics[width=80mm]{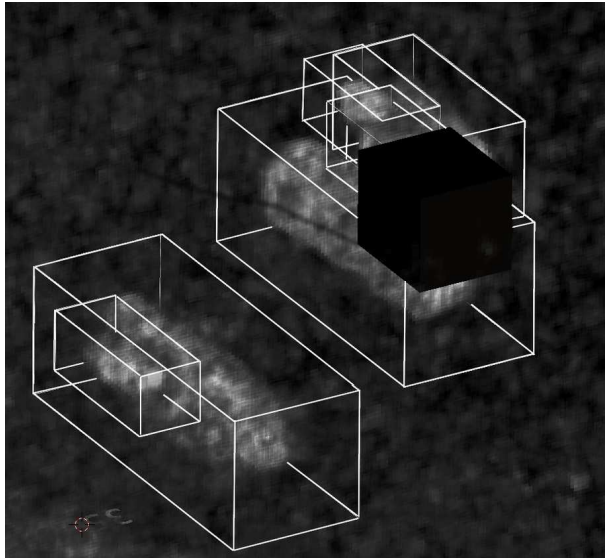}
\caption[Mask]{Masking data in FRELLED, showing wireframe and solid display modes. These modes can be switched interactively. It is possible to create masks of arbitrarily complicated geometry.}
\label{fig:Masks}
\end{center}
\end{figure}

The interactive masking capability is a very important feature of \textsc{Frelled}\footnote{Ordinarily the data is merely hidden from view, however it is also possible to use the masks to actually remove data (i.e. set to zero or NaN).}. The user can click on a source and add an object which is then used to mask the data, scaling this interactively with the mouse or more precisely using numerical input via the keyboard. Masking a source is a procedure which normally takes only a few seconds. This mask object is automatically assigned a name, and the user can easily re-find this object (and re-orient the view appropriately). With hundreds of sources this feature is invaluable. Masks can also be set to be transparent or opaque interactively, and they can be either simple cuboids or have arbitrarily complex geometry. An example of masking is shown in figure \ref{fig:Masks}.

P15 note that a viewer should be capable of fast basic interactions such as pan, zoom, and rotate, at at least 15 fps. \textsc{Frelled} meets or exceeds this requirement even for the largest data sets it can display. A more powerful GPU would enable it to display larger data sets, though as P15 note standard desktop hardware is not likely to dramatically improve on the short term.

In order to reveal different structures the user must be able to interactively alter how the data is displayed. For example, revealing faint sources requires that data is shown with a high contrast at low flux values. This means that bright sources appear saturated, appearing as uniform and featureless. For point sources this is not usually a problem but the difficulties become obvious with extended sources with complex structures.

Choice of the colour transfer function (logarithmic rather than linear) can alleviate this problem, but the best solution is interactive. Blender allows clipping of low transparency (i.e. flux) values, which can help when the data is noisy and contains a mix of faint and bright sources. Transparency can also be altered interactively, which can help boost the visibility of faint sources or structures within bright sources. However the ideal solution of being able to alter the colour transfer function interactively is not currently possible with Blender's display engine. In principle the contrast of the PNG images can be altered, but such changes are not visible in the realtime display. Therefore, only a limited form of interactive data display control is possible. In practice this has not been a severe handicap.

\textsc{Frelled} does not allow for interactively smoothing the data. This is certainly possible for small subsets of the data, but for cubes $\sim$600$^{3}$ voxels such a procedure would take several minutes so could hardly be described as interactive. Again, however, the lack of such a feature has not proven to be a severe problem. Furthermore there are a couple of potential workarounds for this issue, which I discuss in the next two sections.

One issue that cannot be overcome (short of the involvement of a Blender developer) is that there is no way to select a different line of sight integration method - \textsc{frelled} is restricted to Blender's sum of values method.

\subsection{Quantitative analysis}
While the human brain is a very powerful pattern-recognition tool, it is essential to quantify the structures discovered. P15 list several capabilities of \textsc{karma} (in fact the \textsc{karma} program \textit{kvis} was one of the main inspirations to develop \textsc{frelled}), all of which are already at least partially implemented in \textsc{frelled}, and all could potentially be fully implemented :

\noindent \textbf{i)} Display a spectrum. This is already partly possible since \textsc{frelled} can act as a GUI for the \textsc{MIRIAD} task \textit{mbspect}, allowing the user to interactively change the profile of the fitted spectrum as well as which regions are masked. It does not presently allow full control of the mbspect parameters or to adjust the position of the object used to fit the profile based on \textit{mbspect}'s position fitting, though this is certainly possible in principle. Similarly, \textsc{frelled} does not yet support intensity profiles (e.g. Gaussian radial profile fits or simple profile measurements along a spatial slice) but again this is no reason that this cannot be implemented.

\noindent \textbf{ii)} Calculation of statistics in a specified volume. Again this is partially possible. The total flux in a specified volume (of arbitrarily complex shape) can be computed and it would be trivial to add the calculation of the minimum, maximum, rms value etc.

\noindent \textbf{iii)} Segmentation of the 3D data volume of an object, e.g. isosurface fitting. Also partially possible. At present the user can generate renzograms (contour plots at multiple velocities). Converting these into true isosurfaces is literally a matter of joining the dots, for which Blender-based Python code already exists - it is simply a matter of implementation.
By restricting the analysis to a relatively small subset. this feature could also be a way to overcome the lack of an interactive smoothing tool.

\noindent \textbf{iv)} Construction and display of moment maps and position-velocity diagrams. Currently only moment 0 (integrated flux) maps can be produced, but as with the other features there is no reason this could not be extended to other moment maps (e.g. peak flux, velocity maps).

Necessity being the mother of invention, \textsc{frelled}'s development has been concentrated on the features most useful for analysis of AGES (Arecibo Galaxy Environment Survey, \citealt{auld}) data - but there is no reason at all why many more analysis tasks could not be implemented in \textsc{frelled}. Plotting pixel-by-pixel trends in the data and analysing trends in different regions (as described in P15 in relation to the \textsc{GLUE} project of \citealt{good}) would certainly be possible within Blender/\textsc{frelled}.

\subsection{Comparative analysis}
P15 describe the necessity of having linked views of 1D, 2D and 3D and/or to compare models with observations. This is already possible in \textsc{frelled}, though the 1D capabilities are highly limited. The 2D views and 3D views in \textsc{frelled} are linked - objects created in either view are visible and editable in the other. This means, for example, that when the bright sources have been quickly found and masked in the 3D view, the user can switch to the 2D view where those sources have already been masked. They can then proceed to search for fainter sources, which is intrinsically easier in a 2D view, without the distraction of bright sources to overwhelm the display.

Furthermore, the user can add multiple 2D or 3D displays and load different data cubes in each one. The viewpoint orientation is always the same in each display - changes in one are propagated to the rest. This ability to display multiple volumes offers another way around the lack of interactive smoothing - if one accepts the limitation of having to deal with pre-computed smoothed cubes, it is already easily possible to load in a smoothed cube into another display area and then switch between the display of the smoothed and raw cube. It also makes it possible to easily compare models/simulations with the observational data. There is also support for n-body particles, which can be directly overlaid onto the display of a FITS file. At present it is not possible to overlay different FITS cubes (e.g. \textsc{Hi} and CO) though this is possible in principle.

P15 describe the usefulness of interactively modelling rotation curves. Since \textsc{frelled} was initially designed for use with AGES data (mainly unresolved sources) this feature is not yet implemented but again there is really no reason why it could not be.

\subsection{Software interface and code structure}
P15 describe the importance of the quality of the code and accessibility of the software. They note that the software must :

\noindent \textbf{i)} Run on multiple platforms. \textsc{Frelled} was originally developed under Windows 7 and has been run succesfully on Linux (Debian and Ubuntu) and on Mac OS X. In principle there is no reason why it should not run on any system capable of running Blender and Python.

\noindent \textbf{ii)} Have an intuitive interface. I would like to think this is the case ! The underlying drive behind \textsc{frelled}'s GUI was to provide access to Blender's features in a way that astronomers unfamiliar with Blender could easily understand. Thus there are navigation buttons to set the view (e.g. show the XY projection) even though these are not strictly necessary. Auto-import buttons allow the data to be loaded with a single mouse click, and then adjust the settings with more detailed controls later - hopefully this allows a gentler learning curve to visualising 3D data.

\noindent \textbf{iv)} Have a high level of modularity in the source code. Except for a few very similar functions, most operations in \textsc{frelled} call separate subroutines.

\noindent \textbf{v)} Have proper documentation and long-term maintainability. \textsc{Frelled} has a detailed wiki page and manual and well-commented source code. I hope to provide a more detailed guide to the code itself in the near future on the wiki. While it does not yet have a large user or development community, I hope that this report will encourage wider interest. Development of \textsc{frelled} has been almost exclusively by the author and collaborators will be welcomed.

\section{Comparison with Other Viewers}
P15 review four 3D viewers. All of them lack three key features, which I here review in relation to \textsc{frelled} :

\noindent \textbf{i)} Astronomical world coordinate systems. \textsc{Frelled} supports this. For data sets without a WCS (e.g. simulations), pixel values are automatically used. While I do not claim that \textsc{frelled} supports every possible WCS (I thank P15 for graciously providing a data cube with channel velocity units that were previously unsupported) I believe it works with the most common standards. \textsc{Frelled} is usually tested on data cubes from Arecibo (AGES, \citealt{auld}, and GALFA-\textsc{Hi}, \citealt{peek}), the VLA (the VLA GPS survey, \citealt{stil}), Westerbork (the VIRGOHI21 project, \citealt{m07}) and numerical simulation data.

\noindent \textbf{ii)} Interactive smoothing. While, as I have noted, this is not directly possible in \textsc{frelled} (unless we relax our definition of "interactive" to include a process taking several minutes), there are a couple of potential workarounds that may be possible.

\noindent \textbf{iii)} Multi-volume rendering. While \textsc{frelled} also lacks this, in principle it is not difficult to implement. One would simply use alternating image planes to display different data sets with different colour schemes. How well this would work in practise remains to be seen.

\noindent Other features which are sometimes lacking :

\noindent \textbf{iv)} Simple editing or blanking. As described, the interactive masking capabilities of \textsc{frelled} are probably its most useful feature, and one of the key reasons it was developed.

\noindent \textbf{v)} Comparative visualisation. \textsc{Frelled} allows 2D contour plotting, renzograms, on-the-fly overlay of SDSS RGB images, interactive spectral profile measurements, volume-specific flux measurements, and can overlay n-body particles and vectors.

\section{Summary}
\textsc{frelled} is an astronomical 3D data viewer explicitly designed to handle \textsc{Hi} and numerical simulation data. It offers several features that other existing viewers do not. While there are some intrinsic limitations in using Blender as the display engine, in general the advantages far outweigh the disadvantages. Although Blender itself may be ``more suitable for data presentation than analysis'', its Python interpreter completely overcomes this deficiency - any astronomical calculation which can be done in Python can thus be done via, and hence plotted in, Blender. I therefore argue that Blender provides an excellent platform for which to develop advanced 3D data viewing and analysis tools. \textsc{frelled} itself has tremendous scope for additional features. While some of the features described in P15 as desirable are not actually implemented in \textsc{Frelled}, most of them should be straightforward to address.

{}

\end{document}